\documentclass[12pt]{iopart}

\begin{document}

\hspace*{3.2 in}CUQM-117\\
\hspace*{3.5 in}math-ph/0609035\\

\title[Criterion for polynomial solutions $\dots$]{Criterion for polynomial solutions to a class of linear differential equation of second order}

\author{Nasser Saad$^1$, Richard L. Hall$^2,$ and Hakan Ciftci$^3$}

\address{$^1$ Department of Mathematics and Statistics,
University of Prince Edward Island,
550 University Avenue, Charlottetown,
PEI, Canada C1A 4P3.}
\address{$^2$ Department of Mathematics and Statistics, Concordia University,
1455 de Maisonneuve Boulevard West, Montr\'eal,
Qu\'ebec, Canada H3G 1M8}
\address{$^3$ Gazi Universitesi, Fen-Edebiyat Fak\"ultesi, Fizik
B\"ol\"um\"u, 06500 Teknikokullar, Ankara, Turkey.}
\eads{\mailto{nsaad@upei.ca}, \mailto{rhall@mathstat.concordia.ca}, \mailto{hciftci@gazi.edu.tr}}
\def\dbox#1{\hbox{\vrule  
                        \vbox{\hrule \vskip #1
                             \hbox{\hskip #1
                                 \vbox{\hsize=#1}%
                              \hskip #1}%
                         \vskip #1 \hrule}%
                      \vrule}}
\def\qed{\hfill \dbox{0.05true in}}  
\def\square{\dbox{0.02true in}} 
\begin{abstract}
We consider the differential equations $y''=\lambda_0(x)y'+s_0(x)y,$ where $\lambda_0(x), s_0(x)$ are $C^{\infty}-$functions. We prove (i) if the differential equation, has a polynomial solution of degree $n >0$, then $\delta_n=\lambda_n s_{n-1}-\lambda_{n-1}s_n=0,$
where $ \lambda_{n}= \lambda_{n-1}^\prime+s_{n-1}+\lambda_0\lambda_{n-1}\hbox{ and }\quad s_{n}=s_{n-1}^\prime+s_0\lambda_{k-1},\quad n=1,2,\dots.$ Conversely (ii) if $\lambda_n\lambda_{n-1}\ne 0$ and $\delta_n=0$, then the differential equation has a polynomial solution of degree at most $n$.
We show that the classical differential  equations of Laguerre, Hermite, Legendre, Jacobi, Chebyshev (first and second kind), Gegenbauer,  and the Hypergeometric type, etc, obey this criterion. Further, we find the polynomial solutions for the generalized Hermite, Laguerre, Legendre and Chebyshev differential equations. \end{abstract}

\noindent

\submitto{\JPA}
\maketitle

\section{Introduction}
The question as to whether a second order linear homogeneous differential equation has a polynomial solution has attracted much interest since the early classification of Bochner of othogonal polynomials \cite{bo}. In 1929, Bochner posed a problem of determining all families of orthogonal polynomials that are solutions of the differential equation 
\begin{equation}\label{eq1}
(ax^2+bx+c) y''_n(x)+(dx+e)y_n'(x)-\mu_ny_n(x)=0.
\end{equation}
Bochner found that, up to a linear change of variable, only the classical polynomials of Jacobi, Laguerre and Hermite and the Bessel polynomials satisfied a second-order differential equation \cite{gh}-\cite{wa} of the form (\ref{eq1}). In general, the question as to which second order linear homogeneous differential equation has polynomial solutions (not necessary a sequence of orthogonal polynomials) is not easily answered, since it would involve studying a wide variety of equations, including those with regular and irregular singular points. In this article, we present a simple criterion for the existence of polynomial solutions of a differential equation of the form 
\begin{equation}\label{eq2}
y''=\lambda_0 y'+s_0 y
\end{equation}
where $\lambda_0, s_0$ are $C^{\infty}-$functions. 
A key feature of the present work is to note the invariant structure of the right-hand side of (\ref{eq2}) under further differentiation. Indeed, if we differentiate (\ref{eq2}) with respect to $x$, we find that
\begin{equation}\label{eq3}
y^{\prime\prime\prime}=\lambda_1 y^\prime+s_1 y
\end{equation}
where $\lambda_1= \lambda_0^\prime+s_0+\lambda_0^2$ and $s_1=s_0^\prime+s_0\lambda_0.$
If we find the second derivative of equation (\ref{eq3}), we obtain
\begin{equation}\label{eq4}
y^{(4)}=\lambda_2 y^\prime+s_2 y
\end{equation}
where $\lambda_2= \lambda_1^\prime+s_1+\lambda_0\lambda_1\quad\hbox{ and }\quad s_2=s_1^\prime+s_0\lambda_1.$ 
Thus, for $(n+1)^{th}$ and $(n+2)^{th}$ derivative, $n=1,2,\dots$, we have
\begin{equation}\label{eq5}
y^{(n+1)}=\lambda_{n-1}y^\prime+s_{n-1}y
\end{equation}
and
\begin{equation}\label{eq6}
y^{(n+2)}=\lambda_{n}y^\prime+s_{n}y
\end{equation}
respectively, where
\begin{equation}\label{eq7}
\lambda_{n}= \lambda_{n-1}^\prime+s_{n-1}+\lambda_0\lambda_{n-1}\hbox{ ~~and~~ } s_{n}=s_{n-1}^\prime+s_0\lambda_{n-1}.
\end{equation}
From (\ref{eq5}) and (\ref{eq6}) we have
\begin{equation}\label{eq8}
\lambda_n y^{(n+1)}- \lambda_{n-1}y^{(n+2)} = \delta_ny {\rm ~~~where~~~}\delta_n=\lambda_n s_{n-1}-\lambda_{n-1}s_n.
\end{equation}
In an earlier paper \cite{cs} we proved the principal theorem of the Asymptotic Iteration Method (AIM), namely
\vskip0.1in
\noindent{\bf Theorem 1:} \emph{Given $\lambda_0$ and $s_0$ in $C^{\infty}(a,b),$ the differential equation (\ref{eq2}) has the general solution
\begin{equation}\label{eq9}
y(x)= \exp\left(-\int\limits^{x}\alpha(t) dt\right)
\left[C_2 +C_1\int\limits^{x}\exp\left(\int\limits^{t}(\lambda_0(\tau) + 2\alpha(\tau)) d\tau \right)dt\right]
\end{equation}
if for some $n>0$
\begin{equation}\label{eq10}
{s_{n}\over \lambda_{n}}={s_{n-1}\over \lambda_{n-1}} \equiv \alpha.
\end{equation}
}
The present article is not about a classification of orthogonal polynomials which is well-established problem in the literature \cite{bo}-\cite{at}. Rather, the principal goal of the present paper is to characterize when Eq.(\ref{eq2}) has a polynomial solution.  In the next section we shall show that the differential equation (\ref{eq2}) has a polynomial solution of degree $n$, if for some $n>0$,~ $\delta_n = 0.$ In Section~3, we show through detailed analysis that the classical differential  equations of Laguerre, Hermite, Legendre, Jacobi, 
Chebyshev (first and second kind), Gegenbauer,  and the Hypergeometric type, etc., obey this criterion. In Section~4, we apply the criterion to obtain polynomial solutions to the generalized Hermite, Laguerre, Legendre and Chebyshev differential equations. As we shall show, the criterion presented here works whether or not the differential equation (\ref{eq2}) has a set of orthogonal polynomial solutions, or a class of orthogonal polynomial solutions in the quasi-definite sense \cite{ch}.
\section{A criterion for polynomial solutions}
The existence of polynomial solutions is characterized by the vanishing of $\delta_n.$  This is the principal theoretical result of this paper.  We have:
\vskip 0.1in
\noindent{\bf Theorem 2:} {\it (i)~~If the second-order differential equation (\ref{eq2}) has a polynomial solution of degree $n,$ then 
\begin{equation}\label{eq11}
\lambda_n s_{n-1} - \lambda_{n-1} s_n \equiv \delta_n = 0.
\end{equation}
\noindent Conversely~ (ii)~~if $\lambda_n\lambda_{n-1}\ne 0,$ and $\delta_n=0,$ then the differential equation (\ref{eq2}) has a polynomial solution whose degree is at most $n$.}
\vskip0.1in
\noindent{\bf Proof:}~~(i)~~For the given differential equation (\ref{eq2}), if $y$ is a polynomial of degree at most $n$ we have $y^{(n+1)} = y^{(n+2)} = 0.$ consequently we conclude from (\ref{eq8}) that $\delta_n = 0.$
\noindent (ii)~~Conversely, if $\delta_n = 0$ and $\lambda_n\lambda_{n-1}\ne 0,$ then we have $s_{n-1}/\lambda_{n-1} = s_{n}/\lambda_{n}\equiv \alpha,$ and, from Theorem~1, we conclude that a solution is given by $y=\exp(-\int\limits^x \alpha(t) dt).$ Therefore, in particular, $y' = -\alpha y = -{s_{n-1}\over \lambda_{n-1}}y$. Consequently,  from 
$y^{(n+1)}=\lambda_{n-1}y'+s_{n-1}y,$ we infer that $y^{(n+1)}=0,$ or, equivalently, that $y$ is a polynomial of degree at most $n$.\qed
\vskip0.1in
Theorem~2 gives us the condition under which the given differential equation has a polynomial solution. Theorem (1), in particular (\ref{eq9}), provides a tool for the explicit computation of these polynomials. In the next section, we apply these results to a variety of classes of differential equation: in each case we provide the explicit condition which yields polynomial solutions. 
\section{Some differential equations with polynomial solutions}
In this section, we apply Theorem~2 to the classical differential equations of mathematical physics. First, we give an alternative proof to Bochner's results (\ref{eq1}), using the criterion developed in Theorem 2.  
\vskip0.1in
\noindent{\bf Theorem~3:} {\it The second-order differential equation (\ref{eq1}) has a polynomial solution of degree $n$ if 
\begin{equation}\label{eq12}
\mu_n=n(d+(n-1)a),\quad\quad n=0,1,2,\dots.
\end{equation}
The corresponding polynomial solutions are
\begin{eqnarray*}
y_0&=&1\\
y_1&=&dx+e\\
y_2&=&(d+a)(d+2a)x^2+2(b+e)(d+a)x+e(b+e)+c(d+2a)\\
y_3&=&(d+2a)(d+3a)(d+4a)x^3+3(d+2a)(d+3a)(e+2b)x^2\\
&+&3(d+2a)(b(3e+2b)+c(4a+d)+e^2)x\\
&+&4dbc+e^3+3dec+10aec+2eb^2+3e^2b\\
\dots&=&\dots
\end{eqnarray*}
}
\vskip0.1in
\noindent{\bf Proof:} By means of Theorem~2, we find for $\lambda_0=-{dx+e\over ax^2+bx+c}$ and $s_0={\mu\over ax^2+bx+c}$ that the termination condition $\delta_n=\lambda_n s_{n-1}-\lambda_{n-1} s_n =0$ yields
\begin{equation}\label{eq13}
\delta_n=-{1\over (ax^2+bx+c)^{n+1}}\prod_{k=0}^n(k(d+(k-1)a)-\mu_k),\quad n=1,2,\dots
\end{equation}
which yields for $\delta_n=0$ that $\mu_n=n(d+(n-1)a)$ as required. For $n=0,1,2,\dots$ i.e. $\mu_0=0,\mu_1=d, \mu_2=2(d+a),\dots$, we obtian by
\begin{equation}\label{eq14}
y_n=\exp\bigg(-\int\limits^x {s_{n}(t)\over \lambda_{n}(t)}dt\bigg),\quad n=0,1,2,\dots,
\end{equation}
the polynomial solutions just mentioned.\qed
\vskip0.1in

In Table~I, we summarize the well-known differential equations which have polynomial solutions (as eigenfunctions). In each case, we give the explicit criterion, $\delta_n=0$, of Theorem~2. 
\vskip0.2in
\vskip0.1in

\noindent\begin{tabular}{|l|c|c|c|c|}
\hline
DE &  $\lambda_0$&$s_0$&$\delta_n$&$\delta_n=0,\ n=0,1,\dots$\\
\hline
Cauchy-Euler$^1$& ${\alpha(x-b)\over (x-a)^2}$ & ${\beta\over (x-a)^2}$& ${(-1)^{n+1}\over (a-x)^{2n+2}}\prod\limits_{i=1}^{n}(\beta+i(1-i+\alpha))$ &  $\beta=n(n-1-\alpha)$ \\
\hline
Hermite$^{2a}$  & $2x$ & $-2k$& $2^{n+1}\prod\limits_{i=1}^{n}(k-i)$ &  $k=n$ \\
\hline
Hermite$^{2b}$& $ax+b$ & $c$& $(-1)^{n+1}\prod\limits_{i=0}^n (c+ia)$ &  $c=- na$ \\
\hline
Laguerre  & $(1-{1\over x})$ & ${a\over x}$& ${(-1)^{n+1}\over x^{n+1}}\prod\limits_{i=0}^{n} (i+a)$ &  $a=-n$ \\
\hline
Confluent$^3$  & $(b-{c\over x})$ & ${a\over x}$& ${(-1)^{n+1}\over x^{n+1}}\prod\limits_{i=0}^{n} (ib+a)$ &  $a=-nb$ \\
\hline
Hypergeometric  & ${(a+b+1)x-c\over x(1-x)}$ & ${ab\over x(1-x)}$& ${1\over x^{n+1}(x-1)^{n+1}}\prod\limits_{i=0}^{n} (a+i)(b+i)$ &  $a=-n~ or~ b=-n$ \\
\hline
Legendre  & ${2x\over 1-x^2}$ & ${m(m+1)\over x^2-1}$& ${(-1)^n\over (x^2-1)^{n+1}}\prod\limits_{i=0}^{n} {(m^2-i^2)}$ &  $m=n$ \\
\hline
Jacobi  & ${(\alpha+\beta+2)x+\beta+\alpha\over 1-x^2}$ & ${-\gamma\over 1-x^2}$& $\prod\limits_{i=0}^n (i(i+1+\alpha+\beta)-\gamma)$ &  $\gamma=n(n+\alpha+\beta+1)$ \\
\hline 
Chebyshev$^{4a}$& ${x\over 1-x^2}$ & ${-m\over 1-x^2}$& ${(-1)^{n+1}\over (x^2-1)^{n+1}}\prod\limits_{i=0}^{n} (m-i^2)$ &  $m=n^2$ \\
\hline
Chebyshev$^{4b}$& ${3x\over 1-x^2}$ & ${-m\over 1-x^2}$& ${-1\over (x^2-1)^{n+1}}\prod\limits_{i=0}^{n} (i((i+2)-m)$ &  $m=n(n+2)$ \\
\hline
Gegenbauer & ${(1+2k)x\over (1-x^2)}$ & ${-\lambda\over (1-x^2)}$& ${-1\over (x^2-1)^{n+1}}\prod\limits_{i=0}^n(i(i+2k)-\lambda)$ &  $\lambda=n(n+2k)$ \\
\hline
hyperspherical & ${2(1+k)x\over (1-x^2)}$ & ${-\lambda\over (1-x^2)}$& ${-1\over (x^2-1)^{n+1}}\prod\limits_{i=0}^n(i(i+1+2k)-\lambda)$ &  $\lambda=n(n+1+2k)$ \\
\hline
\textmd{Bessel$^{5a}$} & ${-2(x+1)\over x^2}$ & ${\gamma\over x^2}$& ${(-1)^{n+1}\over x^{2n+2}}\prod\limits_{i=0}^n(\gamma-i(i+1))$ &  $\gamma=n(n+1)$ \\
\hline
Generalized& & & & \\
Bessel$^{5b}$& ${-(ax+b)\over x^2}$ & ${\gamma\over x^2}$& ${(-1)^{n+1}\over x^{2n+2}}\prod\limits_{i=0}^n(\gamma-i(i-1+a))$ &  $\gamma=n(n+a-1)$ \\
\hline
\end{tabular}
\vskip0.2in
\noindent{\it {\bf Table~I}: Application of AIM to classical differential equations. For each differential equation which give the condition under which it have polynomial solutions.}
\subsection{Some remarks on Table I}
\begin{enumerate}
\item[$^1$] This differential equation is a generalization of the Cauchy-Euler linear equation 
\begin{equation}\label{eq15}
x^2y''+\alpha xy'+\beta y=0.
\end{equation}
It is possible, however, to apply AIM to the differential equation (\ref{eq15}). The termination condition yields in this case 
\begin{equation}\label{eq16}
\delta_n = {(-1)^{n+1}\over x^{2n+2}}\prod\limits_{i=1}^{n}(\beta+i(1-i+\alpha))=0~\quad\hbox{or}\quad \beta=n(n-1-\alpha)  
\end{equation}
while the corresponding polynomials, as given by (\ref{eq14}), are $y_0=1$, $y_1= x$, $y_2=x^2,\ \dots, y_n=x^n$. It is clear that these polynomials cannot form an orthogonal-polynomial sequence \cite{ch}.
\vskip0.1in
\item[$^{2b}$] This differential equation can be regarded as a generalization of the well-known Hermite differential equation$^{2a}$. It is an elementary example of differential equation with non-rational coefficients (i.e. with $s_0$ and $\lambda_0$ non-rational) which has nonconstant polynomial solutions for $c\neq 0$. 
\vskip0.1in
\item[$^3$] This is known as the confluent hypergeometric differential equation. It is also known as Kummer's differential equation or Pochhammer-Barnes equation \cite{ed}.

\item[$^{4a,b}$] This differential equation is known as Chebyshev's differential equation of the first kind and Chebyshev's differential equation of the second kind, respectively. It is interesting to note that these differential equations are special cases of 
\begin{equation}\label{eq17}
(1-x^2)y''-axy'+\mu y =0.
\end{equation}
If we apply AIM directly to (\ref{eq17}), we have by means of the termination condition (\ref{eq11}) that
\begin{equation*}
\delta_n=-{1\over (x^2-1)^{n+1}}\prod\limits_{k=0}^{n} (i(i+a-1)-\mu_i)
\end{equation*}
thus, for $\delta_n=0$, we must have $\mu_n=n(n+a-1)$. The corresponding polynomial solutions, for $n=0,1,2,\dots$, are
$y_0=1,y_1=x,y_2=(a+1)x^2-1,y_3=(a+1)x^3-3x,\dots,$ and in general
\begin{equation*}
y_n= {}_2F_1(-n,n+a-1,{a\over 2},{1-x\over 2})
\end{equation*}
up to a constant. Here, ${}_2F_1$, Gauss' hypergeometric function, is defined by
\begin{equation}\label{eq18}
{}_2F_1(-n,b;c;x)=\sum\limits_{k=0}^n {(-n)_k (b)_k\over (c)_k k!} x^k,
\end{equation}
where the Pochhammer symbol $(a)_k$ defined by
 \begin{equation*}
(a)_0=1,\quad (a)_k=a(a+1)(a+2)\dots(a+k-1)={\Gamma(a+k)\over \Gamma(a)}.
\end{equation*}
\item[$^{5a,b}$] The polynomial solution of these differential equations were studied by Krall and Frink \cite{kf}. The corresponding (Bessel) polynomial solutions are orthogonal in the quasi-definite sense \cite{ch}.
\end{enumerate}
In Table II we find the corresponding polynomial solutions for each differential equation mentioned in Table I. As an elementary application to quantum mechanics, we consider the Schr\"odinger equation
\begin{equation*}
-{d^2\psi\over dr^2}+\bigg(-{A\over r}+{\gamma(\gamma+1)\over r^2}\bigg)\psi =E\psi.
\end{equation*} 
Writing $\psi(r)=r^{\gamma+1}e^{-\alpha r}y(r)$, we easily find that $y(r)$ must satisfy, for $E=-\alpha^2,$ the confluent hypergeometric differential equation
\begin{equation*}
y''(r)=2\bigg(\alpha-{\gamma+1\over r}\bigg)y'(r)+\bigg({-A+2\alpha(\gamma+1)\over r}\bigg)y(r).
\end{equation*} 
The termination condition, mentioned in Table I, then yields $E=-\alpha^2= -{A^2\over 4(n+\gamma+1)^2}$, the eigenvalues of Schr\"odinger's equation for the Kratzer potential. Furthermore, the corresponding (un-normalized) eigenfunctions are given, by means of Table II, as 
\begin{equation*}
\psi_n(r)=(-1)^nr^{\gamma+1}e^{-\sqrt{-E} r}(2\gamma+2)_n{}_1F_1(-n;2\gamma+2;2\sqrt{-E}r).
\end{equation*}

\noindent{\it {\bf Table~II}: The corresponding polynomial solutions for each differential equation mentioned in Table I.}
\vskip0.1in
\begin{center}
\noindent\begin{tabular}{|l|l|}\hline
DE &  $y_n,~n=0,1,2,\dots$  \\
	\hline
Cauchy-Euler &$y_0= 1$\\ 
~& $y_1=x-b$\\
~&$y_2=(\alpha-1)(\alpha-2)x^2+2(\alpha-1)(2a-\alpha b)x$\\
~~~&~~~~+~$\alpha^2b^2-a(2b+a)\alpha+2a^2$\\
~&$\dots$\\
\hline
Hermite& $y_0(x)=1$\\
~& $y_1(x)=x$\\
~& $y_2(x)=2x^2-1$\\
~&$\dots$\\
~&$y_{2n}(x)=(-1)^n 2^n \left({1/ 2}\right)_n\ {}_1F_1\left(-n;\ {1/ 2};\ x^2\right),$\\
~&$y_{2n+1}(x)=(-1)^n 2^n \left({3/ 2}\right)_n\ x\ {}_1F_1\left(-n;\ {3/ 2};\ x^2\right)$\\
\hline
Hermite& $y_0(x)=1$\\
~&$y_1(x)=ax+b$\\
~&$y_2(x)=(ax+b)^2-a$\\
~&$\dots$\\
~&$y_{2n}(x)=(-1)^n (2a)^n\left({1/2}\right)_n\ {}_1F_1\left(-n;\ {a/2};\ {(ax+b)^2/2}\right),$\\
~&$y_{2n+1}(x)=(-1)^n (2a)^n \left({3/2}\right)_n\ (ax+b)\ {}_1F_1\left(-n;\ {3a/2};\ {(ax+b)^2/2}\right).$\\
\hline
Laguerre  & $y_0=1$\\
~& $y_1=x-1$\\
~& $y_2=x^2-4x+2$\\
~&$\dots$\\
~&$y_n=(-1)^n n!\ {}_1F_1(-n,1,x)$\\
\hline
Confluent&$y_0=1$\\
~& $y_1=bx-c$\\
~&$y_2=(1 + c)c -2 b(1 + c) x + b^2 x^2$\\
~&$\dots$\\
~&$y_n=(-1)^n (c)_n\ {}_1F_1(-n,c,bx)$\\
\hline
Hypergeometric&$y_0=1$\\
~&$y_1=x+c$\\
~&$y_2=2x^2+4(c+1)x+c(c+1)$\\
~&$\dots$\\
~&$y_n=(c)_n~{}_2F_1(-n,-n; c, x)$\\
\hline
\end{tabular}
\end{center}
\begin{center}
\begin{tabular}{|l|l|}
	\hline
DE &  $y_n,~n=0,1,2,\dots$  \\
	\hline
Legendre  & $y_0=1$\\
~&$y_1=x$\\
~&$y_2=-1+x^2$\\
~&$\dots$\\
~&$y_n={}_2F_1(-n,1+n; 1, {(1-x)/2}).$\\
\hline
Jacobi& $y_0=1$\\
~& $y_1=(\alpha-\beta)+(2+\alpha+\beta)x$\\
~& $y_2=(3+\alpha+\beta)(4+\alpha+\beta)x^2+2(\alpha-\beta)(3+\alpha+\beta)x-4-c-d+(c-d)^2$\\
~&$\dots$\\
~&$y_n={(\alpha+1)_n/ n!}~{}_2F_1(-n,n+\alpha+\beta+1;\alpha+1;{(1-x)/ 2}).$\\
\hline
Chebyshev& $y_0=1$\\
~&$y_1=x$\\
~&$y_2=2x^2-1$\\
~&$\dots$\\
~&$y_n={}_2F_1(-n,n,{1\over 2},{(1-x)/2})$\\
 \hline
Chebyshev& $y_0=1$\\
~& $y_1=x$\\
~&$y_2=4x^2-1$\\
~&$\dots$\\
~&$y_n=(n+1){}_2F_1(-n,n+2,{3\over 2},{(1-x)/2})$\\
\hline
Gegenbauer  & $y_0=1$\\
~&$y_1=x$\\
~&$y_2=2(k+1)x^2-1$\\
~&$\dots$\\
~&$y_n={(2k)_n}~{}_2F_1(-n,n+2k;k+{1/2};{(1-x)/2})$\\
\hline
hyperspherical&$y_0=1$\\
~& $y_1=x$\\
~&$y_2=(2k+3)x^2-1$\\
~&$\dots$\\
~&$y_n={(2k+1)_n}~{}_2F_1(-n,n+2k+1;k+1;{(1-x)/2})$\\
\hline
Bessel& $y_1(x)=1+x$\\
polynomials&$y_2(x)=1+3x+3x^2$\\
~&$y_3(x)=1+6x+15x^2+15x^3$\\
~&$\dots$\\
~&$y_n(x)={}_2F_0(-n,n+1;-;-{x/2})$\\
\hline
Generalized Bessel& $y_1(x)=ax+b$\\
polynomials&$y_2(x)=(a+1)(a+2)x^2+2b(a+1)x+b^2$\\
~&$y_3(x)=(a+2)(a+3)(a+4)x^3+3b(a+2)(a+3)x^2+3b^2(2+a)x+b^3$\\
~&$\dots$\\
~&$y_n(x)=b^n{}_2F_0(-n,n+a-1;-;-{x/b})$\\
\hline
\end{tabular}
\end{center}

\subsection{The case of $\lambda_0=0$}
In the early development of the asymptotic iteration method \cite{cs}, one get the impression that the method is not applicable in the case of $\lambda_0=0$. This impression naturally arises  because of the condition ${s_n\over \lambda_n}={s_{n-1}\over \lambda_{n-1}}$, $n=0,1,2,\dots$.  If $\lambda_0=0$, however, we may have using (\ref{eq5}) and (\ref{eq6}) that ${y_{n+2}\over y_{n+1}}={s_n({\lambda_n\over s_n}y'+y)\over s_{n-1} ({\lambda_{n-1}\over s_{n-1}}y'+y)}$
for which the corresponding asymptotic condition now reads
\begin{equation}\label{eq19}
{\lambda_{n}\over s_{n}}={\lambda_{n-1}\over s_{n-1}} \equiv \alpha, \quad\quad n=1,2,\dots
\end{equation} 
This leads to the essentially equivalent termination condition (\ref{eq11})
\begin{equation*}
\delta_n={\lambda_{n} s_{n-1}}-{\lambda_{n-1} s_{n}}=0, \quad\quad n=1,2,\dots
\end{equation*}
A simple example which show the use of AIM in case of $\lambda_0=0$ is the differential equation $x^2y''-2y=0$. Direct use of AIM implies that $\delta_2=0$ and a polynomial solution by mean of (\ref{eq14}) is $y=x^2.$
\section{Application to generalized Hermite, Laguerre, Legendre and Chebyshev differential equations}
\noindent{\bf Theorem~4:}~~{\it For $N$ a positive integer and $a,b\neq 0$, the second-order linear differential equation (known as the generalized Laguerre differential equation)
\begin{equation}\label{eq20}
u''=(ax^N-{b\over x})u'-acx^{N-1}u,
\end{equation}
has a polynomial solution if
\begin{equation}\label{eq21}
c=n(N+1),\quad n=0,1,2,\dots.
\end{equation}
The corresponding polynomial solutions are  
\begin{equation}\label{eq22}
u_n(x)=(N+1)^n\bigg({b+N\over 1+N}\bigg)_n{}_1F_1(-n;{b+N\over 1+N};{ax^{N+1}\over 1+N})
\end{equation}}
\medskip
\noindent{Proof:}~~For $N=1$, the termination condition (\ref{eq11}) yields $c=2n$, $n=0,1,2,\dots$ while (\ref{eq14}) implies
$$u_n(x)=\cases{1,&if $n=0$ (or $c=0$) \cr
	1+b-ax^2,&if $n=1$ (or $c=2$)\cr
	3+4b+b^2-2a(3+b)x^2+a^2x^4,&if $n=2$ (or $c=4$)\cr
\dots\cr
2^n\bigg({b+1\over 2}\bigg)_n{}_1F_1(-n;{b+1\over 2};{ax^2\over 2}),&for $n=0,1,2,\dots$ (or $c=2n$)
\cr}
$$
 For $N=2$, the termination condition (\ref{eq11}) yields $c=3n$, $n=0,1,2,\dots$ while (\ref{eq14})  implies
$$u_n(x)=\cases{1,&if $n=0$ (or $c=0$) \cr
	2+b-ax^3,&if $n=1$ (or $c=3$)\cr
	10+7b+b^2-2a(5+b)x^3+a^2x^6,&if $n=2$ (or $c=6$)\cr
\dots\cr
3^n\bigg({b+2\over 3}\bigg)_n{}_1F_1(-n;{b+2\over 3};{ax^3\over 3}),&for $n=0,1,2,\dots$ (or $c=3n$)
\cr}
$$
 Similarly, for $N=3$, the termination condition (\ref{eq11}) yields $c=4n$, $n=0,1,2,\dots$ while (\ref{eq14}) implies
$$u_n(x)=\cases{1,&if $n=0$ (or $c=0$) \cr
	3+b-ax^4,&if $n=1$ (or $c=4$)\cr
	21+10b+b^2-2a(7+b)x^4+a^2x^8,&if $n=2$ (or $c=8$)\cr
\dots\cr
4^n\bigg({b+3\over 4}\bigg)_n{}_1F_1(-n;{b+3\over 4};{ax^4\over 4}),&for $n=0,1,2,\dots$ (or $c=4n$)
\cr}
$$
Similar expressions can be obtain for $N=4,5,\dots$. These results can be generalized by  (\ref{eq22}). \qed
\vskip0.1in
\noindent{\bf Theorem~5:}~~{\it If $b=0$, the second-order linear differential equation (\ref{eq20}), known as generalized Hermite differential equation, has a polynomial solution if
\begin{equation}\label{eq23}
c=n(N+1),\quad n=0,1,2,\dots
\end{equation}
or
\begin{equation}\label{eq24}
c=n(N+1)+1\quad n=0,1,2,\dots
\end{equation}
In case of $c=n(N+1)$, the polynomial solutions are (for $n=0,1,2,\dots$)
\begin{equation}\label{eq25}
u_n(x)=(-1)^n (N+1)^n\bigg({N\over 1+N}\bigg)_n{}_1F_1(-n;{N\over 1+N};{ax^{N+1}\over 1+N}).
\end{equation}
In case of $c=n(N+1)+1$, the polynomial solutions are (for $n=0,1,2,\dots$)  
\begin{equation}\label{eq26}
u_n(x)=(-1)^n (N+1)^n\bigg({2+N\over 1+N}\bigg)_n~x~{}_1F_1(-n;{2+N\over 1+N};{ax^{N+1}\over 1+N}).
\end{equation}
}
\noindent Proof: Similarly to the proof of Theorem~4, the conditions (\ref{eq23}) and (\ref{eq24}) follow directly by means of the termination condition (\ref{eq11}), with $\lambda_0=ax^N$ and $s_0=-acx^{N-1}.$ Eqs (\ref{eq25}) and (\ref{eq26}) follow from (\ref{eq14}). \qed
\vskip0.1in
\noindent{\bf Theorem~6:}~~{\it For $N$ a positive integer, the differential equation 
\begin{equation}\label{eq27}
u''=\bigg({ax^N\over 1-sx^{N+1}}-{b\over x}\bigg)u'-{wx^{N-1}\over 1-sx^{N+1}}u,
\end{equation}
has polynomial solutions
\begin{eqnarray}\label{eq28}
u_n(x)&=&{(-1)^n\over (N+1)^{-n}}\bigg({N+b\over N+1}\bigg)_n\nonumber\\
&\times& {}_2F_1(-n,{b-1\over N+1}+{a\over (N+1)s}+n;{b+N\over 1+N};sx^{N+1})
\end{eqnarray}
if
\begin{equation}\label{eq29}
w=n(N+1)(s(b-1+n(N+1))+a),\quad n=0,1,2,\dots
\end{equation}
where ${}_2F_1$ is Gauss's hypergeometric function (\ref{eq18}). If $b=0, s=1$, then the cases of $a=2$ and $a=1$ corresponding to differential equations known as the generalized Legendre and Chebyshev differential equations, respectively. 
}
\vskip0.1in
\noindent{\bf Proof:}~~Using AIM, condition (\ref{eq29}) for polynomial solutions follows by means of the termination condition $\delta_n=0$ in a  similar fashion to the proof of Theorem~4. Equation (\ref{eq28}) then follows by means of (\ref{eq14}) as generalization of the polynomial solutions for each of $n=0,1,2,\dots$ and $N=1,2,\dots$. \qed
\vskip0.1in
\section{Conclusion}
We have presented a simple criterion for the existence of polynomial solutions of second-order linear differential equations. Many of the classical differential equations that appear in mathematical physics can be analysed with this theory.  Apart from its theoretical interest, the criterion can be used in a practical way to look for and to obtain polynomial solutions to eigenvalue problems of Schr\"odinger-type \cite{cs}-\cite{bb}, and similarly for polynomial solutions of quasi-exact solvable models in quantum mechanics \cite{sh}. 

\section*{Acknowledgments}
\medskip
\noindent Partial financial support of this work under Grant Nos. GP3438 and GP249507 from the 
Natural Sciences and Engineering Research Council of Canada is gratefully 
acknowledged by two of us (respectively [RLH] and [NS]).

\end{document}